\documentclass[aps,prl,twocolumn]{revtex4-1}
\usepackage{amssymb}
\usepackage{amsmath,bm}
\usepackage{graphicx,color}
\usepackage{bbm}
\usepackage{multirow}
\newcommand{\B}[1]{{\bm{#1}}}

\usepackage{hyperref}

\begin{document}

\title{Universal Density of Low Frequency States in Amorphous Solids at Finite Temperatures}
\author{Prasenjit Das$^1$ and Itamar Procaccia$^{1,2}$}
\affiliation{$^1$Department of Chemical Physics, The Weizmann Institute of Science, Rehovot 76100, Israel. \\$^2$  Center for OPTical IMagery Analysis and Learning, Northwestern Polytechnical University, Xi'an, 710072 China.  }

\begin{abstract}
It has been established that the low frequency quasi-localized modes of amorphous solids at zero temperature exhibit universal density of states, depending on the frequencies as $D(\omega) \sim \omega^4$. It remains an open question whether this universal law extends to finite temperatures. In this Letter
we show that well quenched model glasses at temperatures as high as $T_g/3$ possess the same universal density of states. The only condition required is that
{\em average} particle positions stabilize before thermal diffusion destroys the cage structure of the material.  The universal density of quasi-localized low frequency modes refers then to vibrations around the thermally averaged configuration of the material. 
\end{abstract}

\maketitle

{\bf Introduction}: Amorphous solids exhibit many properties that set them apart from regular elastic solids \cite{11HKLP}. One such aspect attracted high attention for a long time: the density of states of low frequency modes. On purely theoretical grounds it was predicted for more than thirty years now \cite{83KKI,87IKP,91BGGS,03GC,03GPS,07PSG,14SBG} that athermal amorphous solids should exhibit a
density of states $D(\omega)$ with a universal dependence on the frequency $\omega$, i.e.
\begin{equation}
D(\omega)\sim \omega^4 \quad \text{in all dimensions} \  .
\label{dof}
\end{equation}
The actual verification of this theoretical prediction was however slow in coming. The difficulty stemmed from the fact that 
the modes which are expected to exhibit this universal scaling are quasi-localized vibrational modes (QLM) that
in large systems hybridize strongly with low frequency delocalized elastic extended modes. The latter are expected to follow the Debye theory, with density of states depending on frequency like $\omega^{d-1}$ where $d$ is the spatial dimension.

To disentangle these different types of modes it is advantageous to consider small systems. There the (Debye) delocalized modes have a lower cutoff $\omega_{\rm min}\sim 1/L$ (with $L$ the system size), exposing cleanly the QLM with their universal density of states Eq.~(\ref{dof}).  In the recent literature there were a number of direct verifications of this law,
using numerical simulations of glass formers with binary interactions \cite{16LDB,15BMPP,shimada2018spatial,angelani2018probing,17MSI,18KBL}, and also more recently in models of silica glass with binary and ternary interactions \cite{20BGMPZ,20GRKPVBL}. The unusual robustness of this universal
law to extreme changes in the microscopic Hamiltonian was presented and discussed in Ref.~\cite{20DHEP}.

All this progress was achieved in athermal amorphous solids, where the bare Hamiltonian $U(\B r_1, \cdots \B r_N)$  provides the Hessian matrix $\B H$ which determines, in the
harmonic approximation, all the modes and their frequencies
\begin{equation}
H_{ij}^{\alpha \beta} \equiv \frac{\partial^2 U(\B r_1, \cdots \B r_N) }{\partial r_i^\alpha \partial r_j^\beta} \ .
\end{equation}
Here $\B r_i$ is the $i$th coordinate of a constituent particle in a system with $N$ particle. 
As long as the $T=0$ configuration is stable, all the eigenvalues of the
bare Hessian are real and positive (with the exception of few possible zeros associated with Goldstone modes). It is thus straightforward to examine all the modes to determine the density of states. It remains however an open question what is the effect of temperature on the density of states. Before answering this question one needs to ask ``which states?". It is well known that the Hessian computed from the bare Hamiltonian fluctuates due to thermal motion, and its eigenvalues are not bounded by zero from below. Negative eigenvalues (or imaginary frequencies) abound. Thermal motion dresses the bare Hamiltonian; inter-particle collisions impart momentum, giving rise to effective forces that are very different from the bare ones \cite{06BW,16GLPPRR,18PPPRS,19PPSZ}. 
The effective potential from which such renormalized forces can be derived is in general not known. Nevertheless, in Ref.~\cite{19DIP} it was
shown that in thermal glasses in which the {\em average} positions of the particles stabilize (as a time average) before the thermal diffusion
destroys this structure, the effective Hessian associated with vibrations about these average position can be usefully defined. It was demonstrated
that as long as the time-averaged configuration is stable, the eigenvalues of the effective Hessian are again semipositive. An eigenvalue 
approaching zero indicates an incipient instability via saddle-node bifurcation of the time-averaged configuration in much the same way as in athermal conditions \cite{19DIP}.

To define our states, consider a glassy system composed of $N$ particles with time dependent positions $\{\B r_i(t)\}_{i=1}^N$ which is
endowed with a bare Hamiltonian $U\left(\B r_1(t), \cdots, \B r_N(t)\right)$. Assume that the system is in temperature $T$, and that it is
sufficiently stable 
so that the glassy relaxation time (or the effective
diffusion time) $\tau_G$ is long enough, allowing one to compute the time averaged positions $\B R_i$:
\begin{equation}
\B R_i \equiv \frac{1}{\tau}\int_0^\tau dt ~\B r_i(t) \ , \label{defRi}	
\end{equation}
where $\tau\ll \tau_G$.
By definition the position $\B R_i$ are time independent and the configuration $\{\B R_i\}_{i=1}^N$ is stable, at least
for the time interval $[0, \tau_G]$. In addition to the mean positions we need also the covariance matrix $\B \Sigma$ defined
as
\begin{equation}
\B \Sigma_{ij}\equiv \frac{1}{\tau}\int_0^\tau dt \left(\B r_i(t) - \B R_i\right)\left(\B r_j(t) - \B R_j\right) \ .
\label{covariance}
\end{equation}

We can now define an effective Hessian via
\begin{equation}
{\bf H}^{(\rm eff)}=k_B T {\bf \Sigma}^+,
\label{PDhPI}
\end{equation}
where ${\bf \Sigma}^+$ is the pseudo inverse of the covariance matrix.
Of course, the effective Hessian given by Eq.~(\ref{PDhPI})  and the covariance matrix have the same set of eigenfunctions
\begin{equation}
{\bf H}^{(\rm eff)}{\bf \Psi}_i=\lambda^H_i {\bf \Psi}_i
\label{effeig}
\end{equation}
and their eigenvalues are related by
\begin{equation}
\lambda^H_i=\frac{k_B T}{\lambda_i^{\Sigma^+}} \ .
\label{EigV}
\end{equation}
In Ref.~\cite{19DIP} it was shown that the eigenvalues and eigenfunctions of ${\bf H}^{(\rm eff)}$ serve the same role for the time-averaged
configuration as the corresponding ones for the bare Hessian play for the athermal configuration. For example the last nonaffine response to shear
before an instability, matches the eigenfunction of the eigenvalue that vanishes at the instability. It should be noted that this approach was used before to study the vibrational spectra of colloids and granular systems based on measurements using video and  confocal microscopy (see e.g., \cite{10GCDKB,12HBD}),
and restoration of effective interaction potential using simulation methods \cite{16SM}. Our aim here is to study, using numerical simulations of typical
glass formers at finite temperatures,  the density of states of their low-lying modes using this effective Hamiltonian. 

{\bf Simulations}: To measure the density of low frequency modes we employ a standard  model of a glass former, i.e. a binary
mixture of point particles interacting via inverse power-law potentials \cite{99PH}. 50\% of the particles
are ``small" (type $A$) and the other 50\% of the particles are ``large" (type $B$). The interaction
between particle $\alpha$ (being $A$ or $B$) and particle $\beta$ (being $A$ or $B$)
are defined as
\begin{equation}
\phi_{\alpha\beta}(r)= \epsilon \Big(\frac{\sigma_{\alpha\beta}}{r}\Big)^{12}\ .
\end{equation}
Here $\sigma_{AA}=1, \sigma_{AB}=1.2$ and $\sigma_{BB}=1.4$. The interaction potential was cut off (smoothly, with two derivatives) at $4.5 \sigma_{\alpha\beta}$.
It is convenient to introduce reduced units, with $\sigma_{AA}$  being the units of length and
$\epsilon=1$ the unit of energy (with Boltzmann's constant being unity). Simulations were performed using a Monte Carlo method
in an NVT ensemble of $N$ particle, with varying $N$ up to $N=8000$.  The equilibration of the system at a target temperature $T$ was done
in two steps. In the first step the particles were distributed randomly in 
a 3-dimensional box
of size $L^3$ with periodic boundary conditions. $L$ was chosen such that at any temperature the density $\rho=0.76$ \cite{99PH}.  We first equilibrate a system  at a temperature $T = 3$ and
then cool it down in steps of $\Delta T=10^{-4}$ to a target temperature $0<T\le 0.1$, where the upper limit
was chosen since in this system $T_g\approx 0.3$ \cite{99PH}. In the second step the system was stabilized further by performing
$5\times 10^5$ Swap Monte Carlo steps with 80\% of the steps being regular and in 20\% of the steps unlike particles were allowed to exchange position. 
This second step was found to be essential in order to tame the thermal diffusion. Without the second step the average positions of configurations at the higher temperature range ($T \approx 0.1$) did not stabilize sufficiently before diffusion destroyed the cage structures. 

After attaining well quenched configurations at the target temperature, we proceed with $6\times 10^5$ regular Monte Carlo steps where
after every step the coordinates $\{\B r_i(t)\}_{i=1}^N$ are determined. The acceptance rate was chosen to be $30 \%$ at all temperatures. Having
the average positions and the trajectory for every particle, the covariance matrix Eq.~(\ref{covariance}) was evaluated, and its
eigenvalues were computed. 

\begin{table}[h!]
	\begin{center}
		\begin{tabular}{c|c|c}
			Bare Hessian $T=0.01$ & Effective Hessian $T=0.01$ & $T=0$ \\
			\hline
			-1.557 & 0.000 & 0.000  \\ \hline
			-1.308 & 0.000 & 0.000  \\ \hline
			0.000& 0.000 & 0.000  \\ \hline
			0.000 & 0.213 & 0.787  \\ \hline
			0.000 & 0.949 & 0.926 \\ \hline
			0.417 & 1.119 & 1.055\\ \hline
			0.505 & 1.338& 1.151 \\ \hline
			0.755& 1.465 & 1.314  \\ \hline
			0.961& 1.556 & 1.386 \\ \hline
			1.201 & 1.589 & 1.644  \\ \hline
				\label{table1}
		\end{tabular}
	\end{center}
	\caption{First ten lowest eigenvalues of the bare and effective Hessian of randomly selected configurations at $T=0.01$, compared to  eigenvalues of the bare Hessian of a randomly selected configuration at $T=0$. The data
	are shown to three digit accuracy, see text for details.}
\end{table}

{\bf Results}: As eluded already, it is crucial to guarantee that thermal diffusion does not destroy the average configuration during
the time of measurement. A very clear criterion for this is the existence of three zero modes of ${\bf H}^{(\rm eff)}$ whose participation
ratio is unity (cf. Eq.~\ref{defPR} below). Configurations that did not conform with this criterion were excluded from the statistics. Let us examine first very low
temperatures, e.g. $T=0.01$. As can be seen in the first column of Table I, even at this low temperature the bare Hessian
contains negative eigenvalues, excluding it as a useful descriptor of the thermal amorphous solid. The second column shows the first ten
lowest eigenvalues of the effective Hessian of a randomly selected average realization at this temperature. As required, the first three eigenvalues are zero. 
For comparison, we show also eigenvalues of the bare Hessian of a randomly selected athermal configuration. It is interesting to note that
even at this low temperature the eigenvalues differ already in the first digit. 

\begin{figure}
	\includegraphics[width=.30\textwidth]{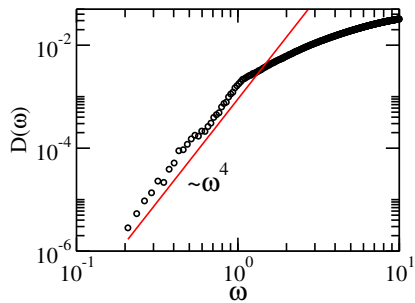}
	\caption{Density of states as computed from the effective Hessian at $T=0.01$. The line with
	slope 4 is a guide to the eye.}
	\label{TDOS0.01} 
\end{figure}

The density of states at this temperature is obtained from
about 5000 independent configurations for each system size (each of which repeats the protocol described above from fresh random initial conditions). In Fig.~\ref{TDOS0.01} we show the result for $N=8000$. The universal scaling law Eq.~(\ref{dof})
is apparent (a least square fit to the data provides a slope of 3.98). 
\begin{figure}
	\includegraphics[width=.30\textwidth]{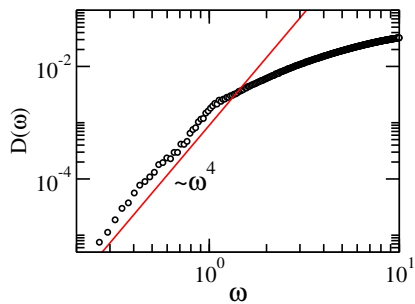}
	\caption{Density of states as computed from the effective Hessian at $T=0.1$. The line with
		slope 4 is a guide to the eye.}
	\label{TDOS0.1} 
\end{figure}

It can be thought that $T=0.01$ is too low for showing the universality of the density of states in thermal systems. So let us consider next simulations 
done at $T=0.1$, which is about a third of $T_g$ \cite{99PH}. With the protocol explained above, the vast majority of our configurations contains three
zero modes, and the few that did not were excluded. The density of states for $N=8000$ is shown in Fig.~\ref{TDOS0.1}.
The universal law (\ref{dof}) appears independent of the temperature as long as the average configuration is stable for sufficiently long time  for
the calculations to converge. 
\begin{figure}
	\includegraphics[width=.30\textwidth]{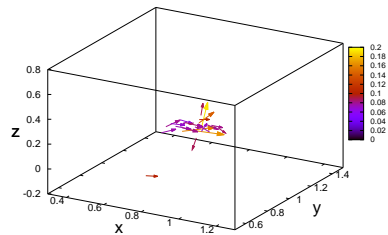}
	\includegraphics[width=.30\textwidth]{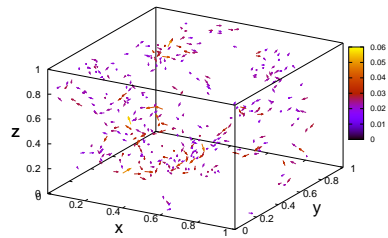}
	\caption{Upper panel: typical QLM from the frequency regime of the universal power law. Its participation ratio is 2.65$\times 10^{-2}$ and its frequency is $\omega=0.56$.  Lower panel: a typical extended mode whose eigen-frequency is above the universal range. Its participation ratio is 0.41 and its frequency is $\omega=3.01$. $T=0.1$ and $N=8000$ in both panels.}
	\label{visuals} 
\end{figure}
To confirm that the modes belonging to the universal regime of $\omega^4$ are indeed quasi-localized we can observe them visually, and we can
compute their participation ratio. A visual is provided in upper panel of Fig~\ref{visuals} where we can see a typical mode whose frequency is in the range of the universal law.
\begin{figure}
	\includegraphics[width=.30\textwidth]{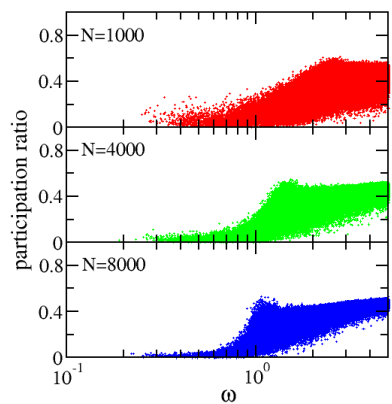}
	\caption{Participation ratio of modes as a function of system size and frequency.}
	\label{PR} 
\end{figure}

To highlight the difference with the modes whose frequency lies above the range of the universal power law we show
such typical mode in the lower panel of Fig~\ref{visuals}. The difference is obvious.  Nevertheless it is worthwhile to
quantify it using the participation ratio $PR$ which is defined as usual \cite{20BGMPZ}
\begin{equation}
PR= [N\sum_i(\B e_i\cdot \B e_i)^2]^{-1} \ ,
\label{defPR}
\end{equation}
where $\B e_i$ is the $i$th element of a given eigenvector of the effective Hessian matrix. We expect the participation ratio
to be of $O(1/N)$ for a QLM and of order unity for an extended mode. The upper panel of Fig.~\ref{visuals} shows a Quasi-Localized mode of frequency 0.56 whose participation ratio is 2.65$\times 10^{-2}$. The lower panel exhibits an extended modes of frequency 3.01 with participation ratio 0.41. The participation ratios as a function of frequency, collected
from all our O(5000) configurations is shown in Fig.~\ref{PR}. We can see that indeed the participation ratio of modes with frequency
$\omega< 1$ is small, $PR\ll 1$, reducing in value with increasing the system size as expected. 

We also learn from this figure that with the increase in system size the delocalized modes whose participation ratio is of $O(1)$ encroach
on lower and lower frequencies. Indeed, for this reason one is limited in system size. With $N=16000$ the regime of the universal power law
is reduced so much that it is hard to discern. Hybridization with extended modes is already there. It should be stressed that the QLM are still 
there, but to expose them one needs to reduce their eigenvalue below the lower cutoff of the extended modes. This can be done by straining
the system to bring it close to a saddle node plastic instability where the eigenvalue approaches zero \cite{12DKP}. There the nonaffine response
of the materials reveals the relevant quasi-localized mode. 

{\bf Summary and Discussion}: In thermal glasses the time averaged positions of the particles are held fixed by renormalized
forces that are determined by momentum transfer. Generically these ``dressed" interaction include binary, ternary, quaternary
and higher order contributions. Moreover, in general the effective Hamiltonian that gives rise to these dressed
interactions is not known analytically. Nevertheless, the effective Hessian of this unknown effective Hamiltonian can be estimated numerically as the pseudo-inverse of the covariance matrix, cf. Eq.~(\ref{PDhPI}). This allows us to determine the density of QLM's in thermal amorphous
solids. The central conclusion of this Letter is that this density conforms with the well studied universal power law Eq.~(\ref{dof}) which
pertained until now to the density of QLM's of the bare Hessian of athermal glasses. This increased applicability of the universal law
seems to indicate that what is important is the amorphous nature of the glassy materials, and not the analytic form of the interaction
potential. It would be fare to say that in spite of all the progress referred to above, a final and convincing theory that
justifies this very broad universality class is still lacking. We trust that the additional evidence presented in this Letter
will add to the urgency of seeking such a theory. 

{\bf acknowledgements}:
This work has been supported in part by the US-Israel Binational Science Foundation and the Joint Laboratory on ``Advanced and Innovative Materials" - Universita' di Roma ``La Sapienza" - WIS.

\bibliography{biblio}
\end{document}